\documentclass[a4,12pt]{article}

\usepackage{cite}

\def \iz {{{}^1\!\!/\!{}_2}}
\def \tz {{{}^3\!\!/\!{}_2}}
\def \pz {{{}^5\!\!/\!{}_2}}

\title{Massless higher spin cubic vertices\\
in flat four dimensional space}

\author{M.V. Khabarov${}^{a}$\thanks{maksim.khabarov@ihep.ru},
Yu.M. Zinoviev${}^{ab}$\thanks{Yurii.Zinoviev@ihep.ru}
\\[0.5cm]
\it{\small ${}^a$Institute for High Energy Physics of National
Research Center "Kurchatov Institute"} \\
\it{\small Protvino, Moscow Region, 142281, Russia} \\
\it\small{ ${}^b$Moscow Institute of Physics and Technology (State
University),} \\
\it{\small Dolgoprudny, Moscow Region, 141701, Russia}}

\date{}

\begin{document}

\maketitle

\begin{abstract}
In this paper we construct a number of cubic interaction vertices
for massless bosonic and fermionic higher spin fields in flat four
dimensional space. First of all, we construct these cubic vertices in
$AdS_4$ space using a so-called Fradkin-Vasiliev approach, which
works only for the non-zero cosmological constant. Then we consider a
flat limit taking care on all the higher derivative terms which
FV-approach generates. We restrict ourselves with the four dimensions
because this allows us to use the frame-like multispinor formalism
which greatly simplifies all calculations and provides a description
for bosons and fermions on equal footing.
\end{abstract}

\thispagestyle{empty}
\newpage
\setcounter{page}{1}

\section{Introduction}

The construction of the cubic interaction vertices for the higher spin
fields is the very first but important step in the investigation of
their consistent interactions. The complete classification of all
cubic vertices for massless and massive bosonic and fermionic fields
were obtained in the light-cone formalism for $d \ge 4$ dimensions by
Metsaev \cite{Met05,Met07b,Met18a}, while the classification for the
massless fields in $d=3$ appeared only quite recently
\cite{Mkr17,KM18}. As for the Lorentz covariant realisation for these
vertices, till now most results deal with the massless fields, where
the main guiding principle is the gauge invariance, which severely
restricts a possible form of the interactions. A lot of interesting
results were developed in the so-called metric-like
formalism (see e.g.
\cite{BBL87,BL06,BFPT06,FT08,Zin08,BLS08,MMR09,BBL10,FT10,
MMR10,MMR10a,MMR10b,JT11,JLT12a,JLT12b,MMR12,JT13,BDT13,CJM16,FMM16,KMP19,
JT19,FKM19} for the bosons and \cite{GHR12,HGR13} for the fermions).
As for the frame-like formalism (which usually leads to the much more
compact and elegant expressions, especially when one uses the
differential form language) the most general results were obtained in
\cite{Vas11} (see also \cite{BPS12}) where the generic cubic vertices
for the massless bosonic fields with spins $s_1\ge s_2 \ge s_3$
satisfying a triangular relation $s_1 < s_2+s_3$ for $AdS_d$ space
with $d \ge 4$  have been constructed. The construction was based on
the so-called Fradkin-Vasiliev approach \cite{FV87,FV87a} where the
non-zero cosmological constant plays a crucial role so that taking a
flat limit appears to be a non-trivial task.

Let us briefly describe the Fradkin-Vasiliev approach to the
construction of cubic vertices. First of all, recall that in the 
frame-like formalism a massless higher spin field is described by the
set of one-forms $\Phi$, each one having its own gauge transformations
(schematically)
$$
\delta_0 \Phi \sim D \eta + e \eta
$$
where $e$ is the background frame. For each one-form a corresponding
gauge invariant two-form (curvature) can be constructed
$$
{\cal R} \sim D \Phi + e \Phi
$$
Moreover, for the non-zero cosmological constant the free Lagrangian
can be rewritten  in the explicitly gauge invariant form
$$
{\cal L}_0 \sim \sum a_k {\cal R}_k {\cal R}_k
$$
where coefficients $a_k$ are determined by the so-called extra field
decoupling conditions. 

The construction of the interactions begins with the most general
quadratic deformations for the initial curvatures
$$
{\cal R} \Rightarrow \hat{\cal R} = {\cal R} + \Delta{\cal R}, \qquad
\Delta {\cal R} \sim \Phi \Phi
$$
One of the nice features of such approach is that these 
deformations simultaneously determine 
the corresponding form for the corrections to the gauge
transformations that can be directly read from that of the curvatures
$$
\delta_1 \Phi \sim \Phi \eta
$$
At this step the main requirement is that these deformed curvatures
must transform covariantly
$$
\delta \hat{\cal R} \sim{\cal R} \eta
$$
Note that the deformation procedure is independent for each of the 
three fields. Then one has to take the sum of the three Lagrangians,
replace initial curvatures by the deformed ones and require that the
resulting Lagrangian be gauge invariant. This leads to the relations
on the previously independent constants and results in  the cubic
vertex that is (on-shell) gauge invariant.

Recall that all cubic vertices can be subdivided into three different
types. The first one we call "trivially gauge invariant" because they
can be written in terms of gauge invariant objects and deform
neither gauge transformations nor gauge algebra. The second type ---
so-called abelian or Chern-Simons like vertices which do have
non-trivial corrections to the gauge transformations, but the algebra
remains to be abelian. At last, the third type --- non-abelian or
Yang-Mills type vertices which deform both the gauge transformations
and the algebra. In \cite{Vas11} Vasiliev has constructed the most
general cubic vertices for the three massless higher spin bosonic
fields in $d \ge 4$ dimensions and shown that they appear to be the
combinations of the non-abelian and abelian vertices, so that all such
vertices from the Metsaev classification \cite{Met05} satisfying the
triangular relation $s_1 < s_2 + s_3$ (assuming $s_1 \ge s_2 \ge s_3$)
are reproduced. Since these vertices have different number of
derivatives, it is not a trivial task to extract a particular vertex
and/or take a flat limit.

The situation is drastically simplified in four dimensions. Indeed, as
has been shown by Metsaev \cite{Met05,Met07b} (see also
\cite{BBL87,CJM16} for bosonic cubic vertices in four dimensions),
all abelian vertices
are absent leaving us only the non-abelian ones. In the frame-like
formalism this result is easy to understand because the abelian
vertices look like ${\cal R}{\cal R}\Phi$ and so must be five-forms.
But even in this case to take the flat limit is not so simple because
the general procedure described above still generate a lot of terms
with a number of derivatives greater than the correct one
($s_1+s_2-s_3$ for bosons and $s_1+s_2-s_3-1$ for fermions). In this
paper we restrict ourselves to the four dimensions and use the
multispinor frame-like formalism (which greatly simplifies all
calculations and allows us to treat bosons and fermions on equal
footing) to reconstruct all non-abelian bosonic and fermionic
vertices. We have managed to show that all these higher derivative
terms combine into total derivatives or cancel on-shell so that we can
safely take a flat limit and obtain (surprisingly) simple form for the
flat vertices. Note that the procedure for the construction of cubic
vertices we use produces only parity even ones, while the results of
\cite{CJM16} show that there exist parity odd vertices with the same
number of derivatives, How these vertices can be reproduced is still
an open question.

The paper is organised as follows. In Section 2 we provide all
necessary information on the multispinor frame-like description for
the massless higher spin bosons and fermions. Sections 3 and 4
contain a number of simple but instructive examples of the vertices
with spin-2 and spin-$\tz$ correspondingly (and, to our opinion, they
are of some interest by themselves). Section 5 contains our results
for the cubic vertices with arbitrary spin bosons and fermions, while
most technical details were moved into two appendices.\\
{\bf Notations ans conventions} We use a formalism where all objects
are multispinors $\Phi^{\alpha(k)\dot\alpha(l)}$, $\alpha,\dot\alpha
= 1,2$ which have $k$ completely symmetric undotted and $l$ completely
symmetric dotted indices. In all expressions where indices are denoted
with the same letter and are placed on the same level, e.g.
$$
\Phi^{\alpha(k)} \Psi^{\alpha(l)}
$$
they are assumed to be symmetrized and symmetrization is defined as
the sum of the minimal number of necessary terms. Besides, all the
fields we consider are the one-forms (and the gauge parameters are
zero-forms), while all the terms in the Lagrangians are the 
four-forms. In this, all the wedge product signs $\wedge$ will be
systematically omitted. 

We work in $AdS_4$ space (and its flat limit) described by the
background frame $e^{\alpha\dot\alpha}$ and the background Lorentz
covariant derivative $D$ satisfying
\begin{equation}
D e^{\alpha\dot\alpha} = 0, \qquad
D D \Phi^{\alpha(k)\dot\alpha(l)} = - \lambda^2
[E^\alpha{}_\beta \Phi^{\alpha(k-1)\beta\dot\alpha(l)} +
E^{\dot\alpha}{}_{\dot\beta} \Phi^{\alpha(k)\dot\alpha(l-1)\dot\beta}]
\end{equation}
where two-forms $E^{\alpha(2)}$ and $E^{\dot\alpha(2)}$ are defined as
follows
\begin{equation}
e^{\alpha\dot\alpha} e^{\beta\dot\beta} = \epsilon^{\alpha\beta}
E^{\dot\alpha\dot\beta} + \epsilon^{\dot\alpha\dot\beta}
E^{\alpha\beta}
\end{equation}

\section{Kinematics}

In this section we provide all necessary information on the frame-like
multispinor formalism for the massless higher spin bosonic and
fermionic fields.

A massless integer spin-$s$ $s > 2$ boson is described by the set of
multispinor one-forms $\Omega^{\alpha(s-1+m)\dot\alpha(s-1-m)}$, 
$0 \le |m| \le s-1$, where $m=0$ corresponds to the physical field,
$m = \pm 1$ --- auxiliary ones, while others are the so-called extra
fields. All fields have their own gauge transformations:
\begin{eqnarray}
\delta \Omega^{\alpha(2s-2)} &=& D \eta^{\alpha(2s-2)} + \lambda^2
e^\alpha{}_{\dot\alpha} \eta^{\alpha(2s-3)\dot\alpha} \nonumber \\
\delta \Omega^{\alpha(s-1+m)\dot\alpha(s-1-m)} &=&
D \eta^{\alpha(s-1+m)\dot\alpha(s-1-m)} + e_\beta{}^{\dot\alpha}
\eta^{\alpha(s-1+m)\beta\dot\alpha(s-2-m)} \nonumber \\
 && + \lambda^2 e^\alpha{}_{\dot\beta} 
\eta^{\alpha(s-2+m)\dot\alpha(s-1-m)\dot\beta} \\
\delta H^{\alpha(s-1)\dot\alpha(s-1)} &=& D 
\eta^{\alpha(s-1)\dot\alpha(s-1)} + e_\beta{}^{\dot\alpha}
\eta^{\alpha(s-1)\beta\dot\alpha(s-2)} + e^\alpha{}_{\dot\beta}
\eta^{\alpha(s-2)\dot\alpha(s-1)\dot\beta}  \nonumber
\end{eqnarray}
Moreover, for each field a gauge invariant two-form can be
constructed:
\begin{eqnarray}
{\cal R}^{\alpha(2s-2)} &=& D \Omega^{\alpha(2s-2)} + \lambda^2
e^\alpha{}_{\dot\alpha} \Omega^{\alpha(2s-3)\dot\alpha} \nonumber 
\\
{\cal R}^{\alpha(s-1+m)\dot\alpha(s-1-m)} &=& D
\Omega^{\alpha(s-1+m)\dot\alpha(s-1-m)} + e_\beta{}^{\dot\alpha}
\Omega^{\alpha(s-1+m)\beta\dot\alpha(s-2-m)} \nonumber \\
 && + \lambda^2 e_\alpha{}^{\dot\beta} 
\Omega^{\alpha(s-2+m)\dot\alpha(s-1-m)\dot\beta} \\
{\cal T}^{\alpha(s-1)\dot\alpha(s-1)} &=& D 
H^{\alpha(s-1)\dot\alpha(s-1)} + e_\beta{}^{\dot\alpha}
\Omega^{\alpha(s-1)\beta\dot\alpha(s-2)} + e^\alpha{}_{\dot\beta}
\Omega^{\alpha(s-1)\dot\alpha(s-1)\dot\beta} \nonumber
\end{eqnarray}
We refer to such two-forms as curvatures. These curvatures satisfy the
following differential identities:
\begin{eqnarray}
D {\cal R}^{\alpha(2s-2)} &=& - \lambda^2 e^\alpha{}_{\dot\alpha}
{\cal R}^{\alpha(2s-3)\dot\alpha} \nonumber \\
D {\cal R}^{\alpha(s-1+m)\dot\alpha(s-1-m)} &=& - 
e_\beta{}^{\dot\alpha} {\cal R}^{\alpha(s-1+m)\beta\dot\alpha(s-2-m)}
- \lambda^2 e_\alpha{}^{\dot\beta} 
{\cal R}^{\alpha(s-2+m)\dot\alpha(s-1-m)\dot\beta} \\
D {\cal T}^{\alpha(s-1)\dot\alpha(s-1)} &=& - e_\beta{}^{\dot\alpha}
{\cal R}^{\alpha(s-1)\beta\dot\alpha(s-2)} - e^\alpha{}_{\dot\beta}
{\cal R}^{\alpha(s-1)\dot\alpha(s-1)\dot\beta} \nonumber 
\end{eqnarray}
On-shell all the curvatures, except the highest ones, are zero,
while the highest one satisfy
\begin{equation}
D {\cal R}^{\alpha(2s-2)} \approx 0, \qquad
e_\beta{}^{\dot\alpha} {\cal R}^{\alpha(2s-3)\beta} \approx 0
\label{idenb}
\end{equation}
Note that zero-curvature conditions imply that on-shell
\begin{eqnarray}
D H^{\alpha(s-1)\dot\alpha(s-1)} &=&  - e_\beta{}^{\dot\alpha}
\Omega^{\alpha(s-1)\beta\dot\alpha(s-2)} - h.c. \nonumber \\
D \Omega^{\alpha(s-1+m)\dot\alpha(s-1-m)} &=&  - 
e_\beta{}^{\dot\alpha} \Omega^{\alpha(s-1+m)\beta\dot\alpha(s-2-m)} +
O(\lambda^2)
\end{eqnarray}
Hence, on-shell the auxiliary field expresses the non-zero derivatives
of the physical field, the extra field
$\Omega^{\alpha(s+1)\dot\alpha(s-3)}$ expresses the non-zero
derivatives of the auxiliary field etc. The field
$\Omega^{\alpha(s-1+m)\dot\alpha(s-1-m)}$ thus expresses the
$m$th derivatives of the physical field which do not vanish on-shell. 
Whenever we talk about the number of derivatives, we imply
the number of derivatives of the physical field and count the $m$th
extra field as an $m$th derivative.

At last, the free Lagrangian can be written in the explicitly gauge
invariant form
\begin{eqnarray}
{\cal L}_0 &=& i(-1)^s \sum_{m=1}^{s-1} 
\frac{(2s-2)!}{(s-1+m)!(s-1-m)!\lambda^{2m}} 
 [{\cal R}_{\alpha(s-1+m)\dot\alpha(s-1-m)} 
{\cal R}^{\alpha(s-1+m)\dot\alpha(s-1-m)} \nonumber \\
 && \qquad \qquad - {\cal R}_{\alpha(s-1-m)\dot\alpha(s-1+m)} 
{\cal R}^{\alpha(s-1-m)\dot\alpha(s-1+m)}]
\end{eqnarray}
Note that the torsion ${\cal T}^{\alpha(s-1)\dot\alpha(s-1)}$ is
absent in this expression. Formally, this Lagrangian contains a lot of
higher derivative terms. However, due to the smart choice of the
coefficients (coming from the so-called extra fields decoupling
conditions) all these terms vanish (up to the total derivatives). So
written in components the Lagrangian reduces to the usual form in
terms of the physical and auxiliary fields only. In particular, it
does not contain any terms singular in the flat limit $\lambda \to 0$.
Recall also that in the multispinor formalism we use parity operation
simply interchanges the dotted and undotted indices and so it
correlates with the conjugation. The choose made (with the imaginary
unit $i$ and minus sign) takes into account that the Lagrangian being
four-form implicitly contains a Levi-Civita symbol.

A massless half-integer spin-$s$ $s > \tz$  fermion is described by a
set of multispinor one-forms $\Phi^{\alpha(s-1+m)\dot\alpha(s-1-m)}$,
$\iz \le |m| \le s-1$, where $m = \pm \iz$ correspond
to the physical fields, all others being the extra ones. The gauge
transformations look very similar to the bosonic case the main
difference is the transformation for the physical fields:
\begin{eqnarray}
\delta \Phi^{\alpha(2s-2)} &=& D \zeta^{\alpha(2s-2)} + \lambda^2
e^\alpha{}_{\dot\alpha} \zeta^{\alpha(2s-3)\dot\alpha} \nonumber \\
\delta \Phi^{\alpha(s-1+m)\dot\alpha(s-1-m)} &=& D
\zeta^{\alpha(s-1+m)\dot\alpha(s-1-m)} + e_\beta{}^{\dot\alpha}
\zeta^{\alpha(s-1+m)\beta\dot\alpha(s-2-m)} \nonumber \\
 && + \lambda^2 e^\alpha{}_{\dot\beta}
\zeta^{\alpha(s-2+m)\dot\alpha(s-1-m)\dot\beta} \\
\delta \Phi^{\alpha(s-\iz)\dot\alpha(s-\tz)} &=& D
\zeta^{\alpha(s-\iz)\dot\alpha(s-\tz)} + e_\beta{}^{\dot\alpha}
\zeta^{\alpha(s-\iz)\beta\dot\alpha(s-\pz)} + \lambda
e^\alpha{}_{\dot\beta} \zeta^{\alpha(s-\tz)\dot\alpha(s-\tz)\dot\beta}
\nonumber
\end{eqnarray}
Similarly, a set of the gauge invariant two-forms can be constructed:
\begin{eqnarray}
{\cal F}^{\alpha(2s-1)} &=& D \Phi^{\alpha(2s-1)} + \lambda^2
e^\alpha{}_{\dot\alpha} \Phi^{\alpha(2s-2)\dot\alpha} \nonumber 
\\
{\cal F}^{\alpha(s-1+m)\dot\alpha(s-1-m)} &=& D
\Phi^{\alpha(s-1+m)\dot\alpha(s-1-m)} + e_\beta{}^{\dot\alpha}
\Phi^{\alpha(s-1+m)\beta\dot\alpha(s-2-m)} \nonumber \\
 && + \lambda^2 e^\alpha{}_{\dot\beta} 
\Phi^{\alpha(s-2+m)\dot\alpha(s-1-m)\dot\beta} \\
{\cal F}^{\alpha(s-\iz)\dot\alpha(s-\tz)} &=& D 
\Phi^{\alpha(s-\iz)\dot\alpha(s-\tz)} + e_\beta{}^{\dot\alpha}
\Phi^{\alpha(s-\iz)\beta\dot\alpha(s-\pz)} + \lambda 
e^\alpha{}_{\dot\beta} \Phi^{\alpha(s-\tz)\dot\alpha(s-\tz)\dot\beta}
\nonumber
\end{eqnarray}
The differential identities for them have the form:
\begin{eqnarray}
D {\cal F}^{\alpha(2s-2)} &=& - \lambda^2 e^\alpha{}_{\dot\alpha}
{\cal F}^{\alpha(2s-3)\dot\alpha} \nonumber \\
D {\cal F}^{\alpha(s-1+m)\dot\alpha(s-1-m)} &=& - 
e_\beta{}^{\dot\alpha} {\cal F}^{\alpha(s-1+m)\beta\dot\alpha(s-2-m)}
- \lambda^2 e^\alpha{}_{\dot\beta}
{\cal F}^{\alpha(s-2+m)\dot\alpha(s-1-m)\dot\beta} \\
D {\cal F}^{\alpha(s-\iz)\dot\alpha(s-\tz)} &=& - 
e_\beta{}^{\dot\alpha} {\cal F}^{\alpha(s-\iz)\beta\dot\alpha(s-\pz)}
- \lambda e^\alpha{}_{\dot\beta}
{\cal F}^{\alpha(s-\tz)\dot\alpha(s-\tz)\dot\beta} \nonumber
\end{eqnarray}
On-shell all these curvatures, except the highest ones, are zero,
while the highest ones satisfy
\begin{equation}
D {\cal F}^{\alpha(2s-2)} \approx 0, \qquad
e_\beta{}^{\dot\alpha} {\cal F}^{\alpha(2s-3)\beta} \approx 0
\label{idenf}
\end{equation}
Again, the zero-curvature conditions imply that the field
$\Phi^{\alpha(s-\iz+m)\dot\alpha(s-\tz-m)}$ expresses 
the $m$th derivatives of the physical field 
$\Phi^{\alpha(s-\iz)\dot\alpha(s-\tz)}$ which do not vanish on-shell.

At last, the free Lagrangian can be written as
\begin{eqnarray}
{\cal L}_0 &=& (-1)^{s+1/2}\sum_{m=\iz}^{s-1} 
\frac{(2s-2)!}{(s-1+m)!(s-1-m)!\lambda^{2m}} \nonumber \\
 && {\cal F}_{\alpha(s-1+m)\dot\alpha(s-1-m)}
{\cal F}^{\alpha(s-1+m)\dot\alpha(s-1-m)} + h.c.
\end{eqnarray}
The same comments on the higher derivative terms, flat limit and
parity as above are applicable here, note however that the absence of
imaginery unit is related with anticommutativity of fermions.

\section{Graviton}

In this section we consider all possible vertices with spin-2 field.
They will serve as the simple illustration for both the general method
and all four possible types of vertices. Besides, interaction with
gravity is always of some interest by itself.

We describe a free massless spin-2 field with the one-forms 
$h^{\alpha\dot\alpha}$, $\omega^{\alpha(2)} + h.c.$ with the initial
gauge transformations
\begin{eqnarray}
\delta \omega^{\alpha(2)} &=& D \eta^{\alpha(2)} - \lambda^2
e^\alpha{}_{\dot\alpha} \xi^{\alpha\dot\alpha} \nonumber \\
\delta h^{\alpha\dot\alpha} &=& D \xi^{\alpha\dot\alpha} +
e_\beta{}^{\dot\alpha} \eta^{\alpha\beta} +
e^\alpha{}_{\dot\beta} \eta^{\dot\alpha\dot\beta}
\end{eqnarray}
The corresponding linearized gauge invariant curvature and torsion
have the form:
\begin{eqnarray}
R^{\alpha(2)} &=& D \omega^{\alpha(2)} + \lambda^2 
e^\alpha{}_{\dot\alpha} h^{\alpha\dot\alpha} \nonumber
\\
T^{\alpha\dot\alpha} &=& D h^{\alpha\dot\alpha} + 
e_\beta{}^{\dot\alpha} \omega^{\alpha\beta} +
e^\alpha{}_{\dot\beta} \omega^{\dot\alpha\dot\beta}
\end{eqnarray}
On-shell we have (note the difference with (\ref{idenb}))
\begin{equation}
T^{\alpha\dot\alpha} \approx 0, \qquad
D R^{\alpha(2)} \approx 0, \qquad
e_\alpha{}^{\dot\alpha} R^{\alpha\beta}
+ e^\alpha{}_{\dot\beta} R^{\dot\alpha\dot\beta} \approx 0
\label{iden2}
\end{equation}
At last, the free Lagrangian can be written as
\begin{equation}
{\cal L}_0 = \frac{i}{\lambda^2} R_{\alpha(2)} R^{\alpha(2)} + h.c.
\end{equation}

There are only two possible types of vertices satisfying the
triangular relation, namely $(s+1,s,2)$ and $(s,s,2)$. For both of
them, the cases with $s=2$ turn out to be special, resulting in four
different cases in total. We consider them in turn.

\subsection{Vertex $(s+1,s,2)$, $s>2$}

We use $\Sigma$ and ${\cal F}$ for the field with spin $s+1$ and its
curvatures and $\Omega$ and ${\cal R}$ --- for spin $s$. Using the
general formulas given in Appendix A, it easy to construct
deformations for the curvatures of all three fields\footnote{Here
and in what follows we provide only the terms which give non-zero
contribution to the flat vertices}. For the spin $s+1$ components we
obtain:
\begin{eqnarray}
\Delta {\cal F}^{\alpha(2s)} &=& a_0\lambda^2 \Omega^{\alpha(2s-2)}
\omega^{\alpha(2)} \nonumber \\
\Delta {\cal F}^{\alpha(2s-1)\dot\alpha} &=& a_0\lambda^2
\Omega^{\alpha(2s-2)} h^{\alpha\dot\alpha} + a_0\lambda^2
\Omega^{\alpha(2s-3)\dot\alpha} \omega^{\alpha(2)} \\
\Delta {\cal F}^{\alpha(2s-2)\dot\alpha(2)} &=& a_0
\Omega^{\alpha(2s-2)} \omega^{\dot\alpha(2)} + O(\lambda^2) \nonumber
\end{eqnarray}
where $a_0$ is a coupling constant and we always choose normalization
so that all coefficients in the deformations are proportional to the
positive degree of $\lambda$. The only variations of the deformed
curvatures that do not vanish on-shell are
\begin{equation}
\delta \hat{\cal F}^{\alpha(2s)} = a_0\lambda^2
\left[ {\cal R}^{\alpha(2s-2)} \eta^{\alpha(2)} - \eta^{\alpha(2s-2)}
R^{\alpha(2)} \right]
\end{equation}
Now we turn to the spin-$s$ components and obtain:
\begin{eqnarray}
\Delta {\cal R}^{\alpha(2s-2)} &=& b_0 \Sigma^{\alpha(2s-2)\beta(2)}
\omega_{\beta(2)} + 2b_0\lambda^2 \Sigma^{\alpha(2s-2)\beta\dot\beta}
h_{\beta\dot\beta} + b_0\lambda^2 \Sigma^{\alpha(2s-2)\dot\beta(2)}
\omega_{\dot\beta(2)} \nonumber \\
\Delta {\cal R}^{\alpha(2s-3)\dot\alpha} &=& b_0
\Sigma^{\alpha(2s-3)\beta(2)\dot\alpha} \omega_{\beta(2)} +
O(\lambda^2)
\end{eqnarray}
In this case, the variations of the deformed curvatures that do not
vanish on-shell are
\begin{equation}
\delta \hat{\cal R}^{\alpha(2s-2)} = b_0 \left[
{\cal F}^{\alpha(2s-2)\beta(2)} \eta_{\beta(2)} - 
\zeta^{\alpha(2s-2)\beta(2)} R_{\beta(2)} \right]
\end{equation}
At last, for the spin-2 we get
\begin{eqnarray}
\Delta R^{\alpha(2)} &=& c_0 \Sigma^{\alpha(2)\beta(2s-2)}
\Omega_{\beta(2s-2)} + (2s-2)c_0\lambda^2
\Sigma^{\alpha(2)\beta(2s-3)\dot\beta} \Omega_{\beta(2s-3)\dot\beta}
\nonumber \\
 && + c_0\lambda^2 \Sigma^{\alpha(2)\dot\beta(2s-2)}
\Omega_{\dot\beta(2s-2)} + O(\lambda^4) \\
\Delta T^{\alpha\dot\alpha} &=& c_0 
\Sigma^{\alpha\beta(2s-2)\dot\alpha} \Omega_{\beta(2s-2)} + c_0
\Sigma^{\alpha\dot\alpha\dot\beta(2s-2)} \Omega_{\dot\beta(2s-2)} +
O(\lambda^2) \nonumber
\end{eqnarray}
with the non-vanishing variations being:
\begin{equation}
\delta \hat{R}^{\alpha(2)} = c_0 \left[{\cal F}^{\alpha(2)\beta(2s-2)}
\eta_{\beta(2s-2)} - \zeta^{\alpha(2)\beta(2s-2)}
{\cal R}_{\beta(2s-2)} \right]
\end{equation}
Now we take the sum of the free Lagrangians and replace the free
curvatures by the deformed ones. The gauge variation of the
resulting Lagrangian produces:
\begin{eqnarray}
\delta \hat{\cal L} &=& \left[ 
\frac{(-1)^{s+1}s(2s-1)a_0}{\lambda^{2s-2}}
+ \frac{(-1)^sb_0}{\lambda^{2s-2}}\right] {\cal
F}_{\alpha(2s-2)\beta(2)}
{\cal R}^{\alpha(2s-2)} \eta^{\beta(2)} \nonumber \\
 && + \left[ \frac{c_0}{\lambda^2} - 
 \frac{(-1)^{s+1}s(2s-1)a_0}{\lambda^{2s-2}}\right] 
{\cal F}_{\alpha(2s-2)\beta(2)} \eta^{\alpha(2s-2)} R^{\beta(2)}
\nonumber \\
 && - \left[ \frac{c_0}{\lambda^2} +
\frac{(-1)^sb_0}{\lambda^{2s-2}}\right]
{\cal R}_{\alpha(2s-2)} \zeta^{\alpha(2s-2)\beta(2)} R_{\beta(2)}
\end{eqnarray}
Thus the invariance of the deformed Lagrangian requires
\begin{equation}
(-1)^{s+1}s(2s-1)a_0 = \lambda^{2s-4}c_0, \qquad
(-1)^sb_0 = - \lambda^{2s-4}c_0
\end{equation}
Now we consider a cubic vertex that follows from the deformed
Lagrangian. Due to the relations on the coupling constants given above
we find that the terms with the highest number of derivatives (and
singular in the flat limit) combine into the total derivative and can
be dropped out. At the next level we obtain terms with the correct
number $N=2s-1$ of derivatives, so we can safely take a flat limit
and, after a number of cancellations, obtain a very simple result:
\begin{equation}
{\cal L}_1 = c_0 D \omega_{\alpha(2)} 
\Sigma^{\alpha(2)\dot\alpha(2s-2)} \Omega_{\dot\alpha(2s-2)} + h.c.
\end{equation}
We see that the spin-2 field enters through the gauge invariant
curvature, while the invariance of the vertex under the other gauge
transformations can be checked using the on-shell identities
(\ref{iden2}) and the corrections to the physical graviton
transformations:
\begin{equation}
\delta h^{\alpha\dot\alpha} = c_0 
\Sigma^{\alpha\dot\alpha\dot\beta(2s-2)} \eta_{\dot\beta(2s-2)} - c_0
\zeta^{\alpha\dot\alpha\dot\beta(2s-2)} \Omega_{\dot\beta(2s-2)} +
h.c.
\end{equation}
%works -> holds
Let us stress that this result holds also for the case when $s$ is
half-integer, i.e. both higher spin fields are fermions.

\subsection{Vertex $(s,s,2)$, $s>2$}

In this case the vertex is symmetric on the two spin-$s$ fields, so
for simplicity we assume that we have just one such field. The part of
the deformation for the spin-$s$ components we need have the form:
\begin{eqnarray}
\Delta {\cal R}^{\alpha(2s-2)} &=& a_0 \Omega^{\alpha(2s-3)\beta}
\omega^\alpha{}_\beta + a_0\lambda^2 \Omega^{\alpha(2s-3)\dot\beta}
h^\alpha{}_{\dot\beta} \nonumber \\
\Delta {\cal R}^{\alpha(2s-3)\dot\alpha} &=& a_0 
\Omega^{\alpha(2s-3)\beta} h^{\dot\alpha}{}_\beta + a_0
\Omega^{\alpha(2s-4)\beta\dot\alpha} \omega^\alpha{}_\beta \nonumber 
\\
 && + a_0 \Omega^{\alpha(2s-3)\dot\beta} 
\omega^{\dot\alpha}{}_{\dot\beta} + O(\lambda^2)
\end{eqnarray}
while the non-vanishing variations of the deformed curvatures look
like:
\begin{equation}
\delta \hat{\cal R}^{\alpha(2s-2)} = a_0 [{\cal R}^{\alpha(2s-3)\beta}
\eta^\alpha{}_\beta - \eta^{\alpha(2s-3)\beta} R^\alpha{}_\beta ]
\end{equation}
The corresponding expressions for the deformations of spin-2 curvature
and torsion are:
\begin{eqnarray}
\Delta R^{\alpha(2)} &=& c_0 \Omega^{\alpha\beta(2s-3)}
\Omega^\alpha{}_{\beta(2s-3)} + c_0\lambda^2
\Omega^{\alpha\beta(2s-4)\dot\beta}
\Omega^\alpha{}_{\beta(2s-4)\dot\beta} \nonumber \\
 && + c_0\lambda^2 \Omega^{\alpha\dot\beta(2s-3)}
\Omega^\alpha{}_{\dot\beta(2s-3)} + O(\lambda^4) \\
\Delta T^{\alpha\dot\alpha} &=& c_0 \Omega^{\alpha\beta(2s-3)}
\Omega^{\dot\alpha}{}_{\beta(2s-3)} + c_0 
\Omega^{\alpha\dot\beta(2s-3)} \Omega^{\dot\alpha}{}_{\dot\beta(2s-3)}
+ O(\lambda^2) \nonumber 
\end{eqnarray}
and for the non-vanishing variations
\begin{equation}
\delta \hat{R}^{\alpha(2)} \sim 2c_0 {\cal R}^{\alpha\beta(2s-3)}
\eta^\alpha{}_{\beta(2s-3)}
\end{equation}
The invariance of the deformed Lagrangian requires
\begin{equation}
(-1)^s(2s-2)a_0 = 4\lambda^{2s-4}c_0
\end{equation}
As in the previous case, due to this relation the terms in  the cubic
vertex with $2s$ derivatives combine into the total derivative and can
be dropped out so that we can safely take a flat limit and obtain
one more simple result:
\begin{equation}
{\cal L}_1 = 2c_0 D \omega_{\alpha\beta} 
\Omega^{\alpha\dot\alpha(2s-3)} \Omega^\beta{}_{\dot\alpha(2s-3)} +
h.c.
\end{equation}
Here the spin-2 also enters only through the gauge invariant
curvature, while the invariance under remaining gauge transformations
holds due to the on-shell identities (\ref{iden2}) and the
corresponding corrections to the physical graviton transformations:
\begin{equation}
\delta h^{\alpha\dot\alpha} = c_0 \Omega^{\alpha\dot\beta(2s-3)}
\eta^{\dot\alpha}{}_{\dot\beta(2s-3)} - c_0
\eta^{\alpha\dot\beta(2s-3)} \Omega^{\dot\alpha}{}_{\dot\beta(2s-3)} +
h.c.
\end{equation}
Note that these results are in agreement with the particular case of
the $(3,3,2)$ vertex which has been considered in \cite{Zin10} (see
also \cite{BL06,Zin08,BLS08} for the metric-like formulation).
Note also that in this case this results works for the fermionic
case where $s$ is half-integer as well.

\subsection{Vertex $(3,2,2)$}

This case is special and provides a simple example of the whole class
of vertices where two lower spins are equal. As far as we know, in
the metric-like formulation this vertex was considered for the first
time in \cite{BL06}, while in the frame-like formalism --- in
\cite{Zin10}. Note that this vertex is antisymmetric on the spin-2
fields so that we must have two different spin-2 particles. 

The deformations for all curvatures have the form now:
\begin{eqnarray}
\Delta {\cal F}^{\alpha(4)} &=& a_0 \Omega^{\alpha(2)}
\omega^{\alpha(2)} \nonumber \\
\Delta {\cal F}^{\alpha(3)\dot\alpha} &=& a_0 \Omega^{\alpha(2)}
h^{\alpha\dot\alpha} + a_0 H^{\alpha\dot\alpha} \omega^{\alpha(2)}
\nonumber \\
\Delta {\cal R}^{\alpha(2)} &=& b_0 \Sigma^{\alpha(2)\beta(2)}
\omega_{\beta(2)} + 2b_0\lambda^2 \Sigma^{\alpha(2)\beta\dot\beta}
h_{\beta\dot\beta} + b_0\lambda^2 H^{\alpha(2)\dot\beta(2)}
\omega_{\dot\beta(2)} \\
\Delta R^{\alpha(2)} &=& c_0 \Sigma^{\alpha(2)\beta(2)}
\Omega_{\beta(2)} + 2c_0\lambda^2 \Sigma^{\alpha(2)\beta\dot\beta}
H_{\beta\dot\beta} + c_0\lambda^2 H^{\alpha(2)\dot\beta(2)}
\Omega_{\dot\beta(2)} \nonumber
\end{eqnarray}
while non-vanishing variations are:
\begin{eqnarray}
\delta \hat{\cal F}^{\alpha(4)} &=& a_0 [{\cal R}^{\alpha(2)}
\eta^{\alpha(2)} -  \zeta^{\alpha(2)} R^{\alpha(2)}] \nonumber \\
\delta \hat{\cal R}^{\alpha(2)} &=& b_0 [
{\cal F}^{\alpha(2)\beta(2)} \eta_{\beta(2)} - 
\zeta^{\alpha(2)\beta(2)} R_{\beta(2)}] \\
\delta \hat{R}^{\alpha(2)} &=& c_0 [
{\cal F}^{\alpha(2)\beta(2)} \zeta_{\beta(2)} - 
\zeta^{\alpha(2)\beta(2)} {\cal R}_{\beta(2)}] \nonumber
\end{eqnarray}
The invariance of the deformed Lagrangian requires
\begin{equation}
6a_0 = \lambda^2b_0, \qquad c_0 = - b_0
\end{equation}
As usual, the terms with 5 derivatives combine into total derivative,
while 3-derivative terms give the following flat vertex:
\begin{equation}
{\cal L}_1 = b_0 D \Omega_{\alpha(2)} H^{\alpha(2)\dot\alpha(2)}
\omega_{\dot\alpha(2)} + 2b_0 \Omega_{\alpha(2)}
\Sigma^{\alpha(2)\beta\dot\alpha} e_\beta{}^{\dot\beta}
\omega_{\dot\alpha\dot\beta} - (\Omega \leftrightarrow \omega) + h.c.
\end{equation}

\subsection{Vertex $(2,2,2)$}

This very well known vertex provides the simplest example of
self-interaction, so for completeness we briefly give it here. The
curvature deformation looks like:
\begin{equation}
\Delta R^{\alpha(2)} = a_0 \omega^{\alpha\beta} \omega^\alpha{}_\beta
+ a_0\lambda^2 h^{\alpha\dot\beta} h^\alpha{}_{\dot\beta}
\end{equation}
The deformed Lagrangian is automatically gauge invariant. The terms
in the cubic vertex with four derivatives combine into the total
derivative leaving us with:
\begin{equation}
{\cal L}_1 = a_0D \omega_{\alpha\beta} h^{\alpha\dot\alpha}
h^\beta{}_{\dot\alpha} - a_0e_\alpha{}^{\dot\alpha} 
h_{\beta\dot\alpha} \omega^{\alpha\gamma} \omega^\beta{}_\gamma + h.c.
\end{equation}

\section{Gravitino}

In this section we present two more simple examples --- vertices with
the spin-$\tz$ field. Taking into account the even in the higher spin
theory the supersymmetry plays a distinguished role, we think they
worth to be considered. The spin-$\tz$ itself is described by the 
one-forms $\psi^\alpha$, $\psi^{\dot\alpha}$ with the gauge invariant
two-forms:
\begin{eqnarray}
F^\alpha &=& D \psi^\alpha + \lambda e^\alpha{}_{\dot\alpha}
\psi^{\dot\alpha} \nonumber \\
F^{\dot\alpha} &=& D \psi^{\dot\alpha} + \lambda 
e_\alpha{}^{\dot\alpha} \psi^\alpha
\end{eqnarray}
and the free Lagrangian
\begin{equation}
{\cal L}_0 = \frac{1}{\lambda} F_\alpha F^\alpha + h.c.
\end{equation}
There are two types of vertices satisfying the strict triangle
inequality and corresponding to the two types of the massless
supermultiplets --- $(s+\iz,s,\tz)$ and $(s+1,s+\iz,\tz)$.

\subsection{Vertex $(s+\iz,s,\tz)$, $s \ge 2$}

We begin with the deformations for all curvatures (keeping only
necessary terms):
\begin{eqnarray}
\Delta {\cal F}^{\alpha(2s-1)} &=& a_0\lambda \Omega^{\alpha(2s-2)} 
\psi^\alpha \nonumber \\
\Delta {\cal F}^{\alpha(2s-2)\dot\alpha} &=& a_0 \Omega^{\alpha(2s-2)}
\psi^{\dot\alpha} + O(\lambda) \nonumber \\
\Delta {\cal R}^{\alpha(2s-2)} &=& b_0 \Phi^{\alpha(2s-2)\beta}
\psi_\beta + b_0\lambda \Phi^{\alpha(2s-2)\dot\beta} \psi_{\dot\beta}
\\
\Delta F^\alpha &=& c_0 \Phi^{\alpha\beta(2s-2)} \Omega_{\beta(2s-2)}
 + c_0\lambda \Phi^{\alpha\dot\beta(2s-2)} \Omega_{\dot\beta(2s-2)}
+ O(\lambda^2) \nonumber 
\end{eqnarray}
Non-vanishing variations have the form:
\begin{eqnarray}
\delta \hat{\cal F}^{\alpha(2s-1)} &=& a_0\lambda 
[ {\cal R}^{\alpha(2s-2)} \zeta^\alpha - \eta^{\alpha(2s-2)} F^\alpha]
\nonumber \\
\delta \hat{\cal R}^{\alpha(2s-2)} &=& b_0 
[ {\cal F}^{\alpha(2s-2)\beta} \zeta_\beta - \zeta^{\alpha(2s-2)\beta}
F_\beta ] \\
\delta \hat{F}^\alpha &=& c_0 [ {\cal F}^{\alpha\beta(2s-2)}
\eta_{\beta(2s-2)} - \zeta^{\alpha\beta(2s-2)} {\cal R}_{\beta(2s-2)}
] \nonumber
\end{eqnarray}
The invariance of the deformed Lagrangian requires
\begin{equation}
(-1)^{s+1}(2s-1)a_0 = \lambda^{2s-3}c_0, \qquad
(-1)^sb_0 = \lambda^{2s-3}c_0
\end{equation}
The resulting flat vertex with the correct number of derivatives 
$N = 2s-2$ (after the higher derivative terms combine
into total derivative and were dropped out) takes the form:
\begin{equation}
{\cal L}_1 = c_0 D \psi_\alpha \Phi^{\alpha\dot\alpha(2s-2)}
\Omega_{\dot\alpha(2s-2)} + h.c.
\end{equation}
Once again we find that the lowest spin field enters through its gauge
invariant curvature only, while to check the invariance under the
remaining gauge transformations one has to take into account the 
corrections to the gravitino gauge transformations:
\begin{equation}
\delta \psi^{\dot\alpha} = c_0 \Phi^{\dot\alpha\dot\beta(2s-2)}
\eta_{\dot\beta(2s-2)} - c_0 \zeta^{\dot\alpha\dot\beta(2s-2)}
\Omega_{\dot\beta(2s-2)} + h.c.
\end{equation}

\subsection{Vertex $(s+1,s+\iz,\tz)$, $s\ge 2$}

This case appears to be very similar, so we will be brief. The
appropriate deformations look like:
\begin{eqnarray}
\Delta {\cal R}^{\alpha(2s)} &=& a_0\lambda \Phi^{\alpha(2s-1)} 
\psi^\alpha  \nonumber \\
\Delta {\cal R}^{\alpha(2s-1)\dot\alpha} &=& a_0 \Phi^{\alpha(2s-1)}
\psi^{\dot\alpha} + O(\lambda) \nonumber \\
\Delta {\cal F}^{\alpha(2s-1)} &=& b_0 \Omega^{\alpha(2s-1)\beta}
\psi_\beta + b_0\lambda \Omega^{\alpha(2s-1)\dot\beta}
\psi_{\dot\beta} \\
\Delta F^\alpha &=& c_0 \Omega^{\alpha\beta(2s-1)} \Phi_{\beta(2s-1)}
 + c_0\lambda \Omega^{\alpha\dot\beta(2s-1)} \Phi_{\dot\beta(2s-1)}
\nonumber 
\end{eqnarray}
while the relations on the coupling constants are:
\begin{equation}
(-1)^{s+1}2sa_0 = - \lambda^{2s-2}c_0, \qquad
(-1)^{s+1}b_0 = \lambda^{2s-2}c_0
\end{equation}
The resulting flat cubic vertex with $N=2s-1$ derivatives appears to
be
\begin{equation}
{\cal L}_1 = c_0 D \psi_\alpha \Omega^{\alpha\dot\alpha(2s-1)}
\Phi_{\dot\alpha(2s-1)} + h.c.
\end{equation}

The results given above hold only for $s \ge 2$, while the case $s=1$
turns out to be special (as all cases where two lowest spins are
equal). This vertex $(2,\tz,\tz)$ is very well known being a part of
the $N=1$ supergravity, but for completeness we briefly provide this
vertex in our current formalism.

The deformations now are very simple
\begin{eqnarray}
\Delta R^{\alpha(2)} &=& \frac{i}{4}c_0\lambda \psi^\alpha \psi^\alpha
\nonumber \\
\Delta T^{\alpha\dot\alpha} &=& \frac{i}{2}c_0 \psi^\alpha 
\psi^{\dot\alpha} \\
\Delta F^\alpha &=& c_0 \omega^{\alpha\beta} \psi_\beta + c_0\lambda
h^{\alpha\dot\alpha} \psi_{\dot\alpha} \nonumber
\end{eqnarray}
and the flat vertex has the form:
\begin{equation}
{\cal L}_1 = c_0 D \psi_\alpha h^{\alpha\dot\alpha} \psi_{\dot\alpha}
- c_0 e_\alpha{}^{\dot\alpha} \psi_{\dot\alpha} \omega^{\alpha\beta}
\psi_\beta + h.c.
\end{equation}

\section{Arbitrary spins}

In  this section we consider general case of three arbitrary spins
$s_1 \ge s_2 \ge s_3$. We introduce their convenient combinations:
\begin{equation}
\hat{s}_1 = s_2 + s_3 - s_1 - 1, \quad
\hat{s}_2 = s_1 + s_3 - s_2 - 1, \quad
\hat{s}_3 = s_1 + s_2 - s_3 - 1
\end{equation}
Note that if spins $s_{1,2,3}$ satisfy the triangular relations
these combinations are always non-negative: $\hat{s}_{1,2,3} \ge 0$.
Moreover, even if two of the three fields are fermions and two of the
three $s_{1,2,3}$ are half-integer, the corresponding 
$\hat{s}_{1,2,3}$ are always integer. Let us give here some useful
relations on them:
\begin{equation}
\hat{s}_1 + \hat{s}_2 = 2(s_3-1), \qquad
\hat{s}_1 + \hat{s}_3 = 2(s_2-1), \qquad
\hat{s}_2 + \hat{s}_3 = 2(s_1-1)
\end{equation}

We begin with the bosonic case and then make necessary adjustment for
the fermionic one. We use notations $\Sigma$, ${\cal F}$ for the
fields component and curvatures for the highest spin $s_1$,
$\Omega$, ${\cal R}$ for spin $s_2$ and $\omega$, $R$ for the lowest
spin $s_3$ correspondingly.

The deformations for all curvatures of the highest spin $s_1$ have the
form:
\begin{equation}
\Delta {\cal F}^{\alpha(2s_1-2-m)\dot\alpha(m)} = 
\sum_{k=0}^{\hat{s}_1} \sum_{l=0}^{\min(m,\hat{s}_2)} a_k 
\Omega^{\alpha(\hat{s}_3-m+l)\beta(\hat{s}_1-k)\dot\alpha(m-l)\dot\beta(k)}
\omega^{\alpha(\hat{s}_2-l)\dot\alpha(l)}{}_{\beta(\hat{s}_1-k)\dot\beta(k)} \label{anz}
\end{equation}
where coefficients $a_k$ (see Appendix A for details) look like
\begin{equation}
a_k = \frac{(\hat{s}_1)!}{(\hat{s}_1-k)!k!}a_0
\end{equation}
Strictly speaking, these coefficients must be multiplied by $\lambda$
raised to some positive power, but to simplify formulas we temporarily
set $\lambda=1$. We restore them by dimensionality of terms whenever
it is necessary. 

Similarly, for the two other spins $s_{2,3}$ we consider
\begin{equation}
\Delta {\cal R}^{\alpha(2s_2-2-m)\dot\alpha(m)} =
\sum_{k=0}^{\hat{s}_2} \sum_{l=0}^{min(m,\hat{s}_1)} b_k
\Sigma^{\alpha(\hat{s}_3-m+l)\beta(\hat{s}_2-k)\dot\alpha(m-l)
\dot\beta(k)} 
\omega^{\alpha(\hat{s}_1-l)\dot\alpha(l)}{}_{\beta(\hat{s}_2-k)
\dot\beta(k)} 
\end{equation}
\begin{equation}
\Delta R^{\alpha(2s_3-2-m)\dot\alpha(m)} =
\sum_{k=0}^{\hat{s}_3} \sum_{l=0}^{\min(m,\hat{s}_1)} c_k
\Sigma^{\alpha(\hat{s}_2-m+l)\beta(\hat{s}_3-k)\dot\alpha(m-l)
\dot\beta(k)} 
\Omega^{\alpha(\hat{s}_1-l)\dot\alpha(l)}{}_{\beta(\hat{s}_3-k)
\dot\beta(k)} 
\end{equation}
with the corresponding coefficients
\begin{equation}
b_k = \frac{(\hat{s}_2)!}{(\hat{s}_2-k)!k!}b_0, \qquad
c_k = \frac{(\hat{s}_3)!}{(\hat{s}_3-k)!k!}c_0
\end{equation}
Now we take a sum of the three Lagrangians, replace the initial
curvatures by the deformed ones and require the resulting deformed
Lagrangian to be invariant. The non-vanishing on-shell variations have
the form:
\begin{eqnarray}
\delta \hat{\cal F}^{\alpha(2s_1-2)} &=& a_0 [ 
{\cal R}^{\alpha(\hat{s}_3)\beta(\hat{s}_1)} 
\eta^{\alpha(\hat{s}_2)}{}_{\beta(\hat{s}_1)} - 
\eta^{\alpha(\hat{s}_3)\beta(\hat{s}_1)} 
R^{\alpha(\hat{s}_2)}{}_{\beta(\hat{s}_1)} ] \nonumber \\
\delta \hat{\cal R}^{\alpha(2s_2-2)} &=& b_0 [ 
{\cal F}^{\alpha(\hat{s}_3)\beta(\hat{s}_2)} 
\eta^{\alpha(\hat{s}_1)}{}_{\beta(\hat{s}_2)} - 
\eta^{\alpha(\hat{s}_3)\beta(\hat{s}_2)} 
R^{\alpha(\hat{s}_1)}{}_{\beta(\hat{s}_2)} ]  \\
\delta \hat{R}^{\alpha(2s_3-2)} &=& c_0 [ 
{\cal R}^{\alpha(\hat{s}_2)\beta(\hat{s}_3)} 
\eta^{\alpha(\hat{s}_1)}{}_{\beta(\hat{s}_3)} - 
\eta^{\alpha(\hat{s}_2)\beta(\hat{s}_3)} 
R^{\alpha(\hat{s}_1)}{}_{\beta(\hat{s}_3)} ] \nonumber 
\end{eqnarray}
Then the invariance of the Lagrangian requires (for what follows it is
important to restore the $\lambda$ dependence here):
\begin{equation}
(-1)^{s_1} \frac{(\hat{s}_2+\hat{s}_3)!}{(\hat{s}_2)!(\hat{s}_3)!}
\frac{a_0}{\lambda^{2s_1-2}}
= - (-1)^{s_2}
\frac{(\hat{s}_1+\hat{s}_3)!}{(\hat{s}_1)!(\hat{s}_3)!}\frac{b_0}{\lambda^{2s_2-2}} =
(-1)^{s_3} \frac{(\hat{s}_1+\hat{s}_2)!}{(\hat{s}_1)!
(\hat{s}_2)!}\frac{c_0}{\lambda^{2s_3-2}} \label{abc}
\end{equation}

Now let us turn to the cubic vertex. Recall, that all the curvatures
except the highest ones, i.e. ${\cal F}^{\alpha(2s_1-2)}$, ${\cal
R}^{\alpha(2s_2-2)}$ and $R^{\alpha(2s_3-2)}$ (and their conjugates),
vanish on-shell. So it seems that the simplest way to obtain the cubic
vertex is to take into account their deformations only. But this
produce a lot of terms with the number of derivatives greater than
$N=s_1+s_2-s_3$, moreover, their coefficients will be proportional to
the negative degrees of $\lambda$ and so will be singular in the flat
limit. Note that due to relation on the constants given above the
terms with the highest number of derivatives, namely $s_1+s_2+s_3-2$
combine into total derivative and can be dropped out, but it still
leaves a lot of other dangerous terms (exceptions are the vertices
with lowest spin-2 and spin-$\tz$). So before taking a flat limit we
must show that all these terms somehow vanish on-shell. It turns out
that the best strategy is to keep all the curvatures and all their
deformations. In this way we managed to show (see Appendix B for
details) that all such terms combine into total derivatives or cancel
each other so we safely can take a flat limit. 
The procedure we followed produce also a lot of terms which have the
correct number $s_1+s_2-s_3$ of derivatives and contribute to the flat
vertex. By rather long but straightforward calculations (ones again
see Appendix B) we reduced the final results to (we dare say) the
simplest form possible.

Among all cubic vertices there are four possible types, namely
$s_1>s_2>s_3$, $s_1=s_2>s_3$, $s_1>s_2=s_3$ and $s_1=s_2=s_3$, and, as
we have seen on the simple examples above, have to be considered
separately.

\subsection{Vertex $s_1>s_2>s_3$}

First of all note that the relation (\ref{abc}) implies that
$$
a_0 \sim \lambda^{2(s_1-s_3)}c_0, \qquad
b_0 \sim \lambda^{2(s_2-s_3)}c_0
$$
It means that in the flat limit all deformations for the two higher
spins vanish and as a result the flat vertex must be trivially
invariant under the lowest spin field gauge transformations. And
indeed, we managed to reduce this vertex to very simple form
\begin{equation}
{\cal L}_1 = 2c_0 D \omega_{\alpha(\hat{s}_2)\beta(\hat{s}_1)}
\Sigma^{\alpha(\hat{s}_2)\dot\alpha(\hat{s}_3)}
\Omega^{\beta(\hat{s}_1)}{}_{\dot\alpha(\hat{s}_3)} + h.c.
\end{equation}
where the lowest spin field enters through the gauge invariant
curvature. As for the invariance under the other gauge
transformations, it can be easily checked with the help of on-shell
identities (\ref{idenb}) or (\ref{idenf}). Recall that even if the two
of the three fields are fermions so that two of the three $s_{1,2,3}$
are half-integer, the combinations $\hat{s}_{1,2,3}$ are always
integer and so the formula above works for the fermionic vertices as
well.

\subsection{Vertex $s_1=s_2>s_3$}

First of all note that these vertices are symmetric on the two higher
spin fields if $s_3$ is even (so it may be one and the same field) and
antisymmetric if $s_3$ is odd. In all other respects, including
considerations on the gauge invariance, they are very similar to the
previous case. The flat vertex turns out to be
\begin{equation}
{\cal L}_1 = 2c_0 D \omega_{\alpha(s_3-1)\beta(s_3-1)}
[\Sigma^{\alpha(s_3-1)\dot\alpha(\hat{s}_3)}
\Omega^{\beta(s_3-1)}{}_{\dot\alpha(\hat{s}_3)} + (-1)^{s_3}
(\Sigma \leftrightarrow \Omega)] + h.c.
\end{equation}
so the lowest spin field also enters only through the gauge invariant
curvature. Note also, that in this case the two higher spin fields can
be fermions, but lower spin field is always boson.

\subsection{Vertex $s_1>s_2=s_3$}

For the even highest spin $s_1$ such vertex must be symmetric on the
two lower spin ones, so it may be one and the same field, while for
the odd $s_1$ it must be antisymmetric and we must have two different
fields with the same spin. We have seen on the simple examples above
that this case is indeed special and the vertex has a more complicated
form. Indeed, the relations on the coupling constants
$$
a_0 \sim \lambda^{2(s_1-s_3)}c_0, \qquad b_0 \sim c_0
$$
show that only corrections to the higher spin transformations vanish
in the flat limit and so the vertex cannot be trivially gauge
invariant under the gauge transformations of the lower spin fields.
The most simple result we have managed to obtain looks like:
\begin{eqnarray}
{\cal L}_1 &=& c_0 D \omega_{\alpha(s_1-1)\beta(\hat{s}_1)}
H^{\alpha(s_1-1)\dot\beta(s_1-1)}
 \Omega^{\alpha(\hat{s}_1)}{}_{\dot\beta(s_1-1)} \nonumber \\
 && + c_0\sum_{k=0}^{\hat{s}_1}
\frac{(s_1-1)(\hat{s}_1)!}{(\hat{s}_1-k)!k!}
e^\gamma{}_{\dot\gamma} \Sigma_{\alpha(s_1-1)\gamma\dot\alpha(s_1-2)}
\Omega^{\alpha(s_1-1)\beta(\hat{s}_1-k)\dot\beta(k)}
\omega^{\alpha(s_1-2)\dot\gamma}{}_{\beta(\hat{s}_1-k)\dot\beta(k)}
\nonumber \\
 && + (-1)^{s_1} (\Omega \leftrightarrow \omega) + h.c
\end{eqnarray}
The first term has the same structure as in the general case the main
difference is that the highest spin enters through its physical
component that has different on-shell relations. As a result, the
first term is not gauge invariant by itself and the gauge invariance
requires that the number of algebraic terms to be added.

Note that in this case the two lower spin fields can be fermions,
while the highest spin one is always boson.

\subsection{Vertex $s_1=s_2=s_3=s$}

Similarly to the previous case, for the even spin $s$ this vertex must
be completely symmetric on all three fields so that it may be just one
and the same field and the vertex describes its self interaction; for
the odd spin $s$ the vertex must be completely antisymmetric so we
must have three different fields with the same spin. In this case
$$
a_0 \sim b_0 \sim c_0
$$
so that the corrections to the gauge transformations for all three
fields survive in the flat limit and the resulting vertex looks very
similar to the previous one:
\begin{eqnarray}
{\cal L}_1 &=& c_0 D \Sigma_{\alpha(s-1)\beta(s-1)}
\Phi^{\alpha(s-1)\dot\alpha(s-1)} 
\phi^{\beta(s-1)}{}_{\dot\alpha(s-1)} \nonumber \\
 && + c_0 \sum_{k=1}^{s-1} 
\frac{(s-1)(s-1)!}{(s-1-k)!k!} e^\gamma{}_{\dot\gamma}
\Sigma_{\alpha(s-1)\gamma\dot\alpha(s-2)}
\Omega^{\alpha(s-1)\beta(s-1-k)\dot\beta(k)}
\omega^{\alpha(s-2)\dot\gamma}{}_{\beta(s-1-k)\dot\beta(k)} \nonumber
\\
 && + min.\ perm.(\Sigma,\Omega,\omega) + h.c.
\end{eqnarray}
Here $ min.\ perm.$ stands for the two cyclic permutations 
of $\Sigma,\Omega,\omega$ in the first term and five permutations
in the second one. It is clear that such vertices exist only for
bosons. 

\section*{Conclusion}

In this paper we have constructed a number of non-trivial cubic
vertices for the massless higher spin bosonic and fermionic fields in
flat four dimensional space. We begin with Fradkin-Vasiliev approach
in $AdS_4$ space and then consider the flat limit. The procedure
appears to be not so simple, because we have to take care on all the
higher derivative terms, which such approach generates, but the final
results happen to be very simple. So we hope that they could be useful
for the future investigations. Let us stress once more that the 
procedure we use produce only parity even vertices, while the
construction of the corresponding parity odd ones \cite{CJM16} is an
open question.

As one of the future directions we see a construction of the
cubic vertices for massive and partially massless fields. The
frame-like formalism for such fields is known
\cite{Zin08b,PV10,KhZ19}, but there are just a few examples of
interactions till now \cite{Zin09,Zin10a,BSZ11,Zin11,Zin14,GMS20}.

One more interesting direction is the cubic vertices for the higher
spin massless supermultiplets. Their classification was elaborated 
quite recently in the light-cone formalism \cite{Met19a,Met19b}, but
for the Lorentz covariant realization there are also just a few
non-trivial results \cite{BGK17,BGK18,BGK18a,BGK18b,GK19}.

\section*{Acknowledgements}
M.Kh. is grateful to Foundation for the Advancement of Theoretical
Physics and Mathematics "BASIS" for their support of the work.

\appendix

\section{Deformations}

In this appendix we calculate the combinatoric coefficients for the
deformations of the gauge invariant curvatures. Let us take as an
example the curvatures of the highest spin components. Their most
general quadratic deformations are given by ansatz (\ref{anz}), which
we repeat here for the reader convenience:
$$
\Delta {\cal F}^{\alpha(2s_1-2-m)\dot\alpha(m)} = 
\sum_{k=0}^{\hat{s}_1} \sum_{l=0}^{\min(m,\hat{s}_2)} a_k 
\Omega^{\alpha(\hat{s}_3-m+l)\beta(\hat{s}_1-k)\dot\alpha(m-l)\dot\beta(k)}
\omega^{\alpha(\hat{s}_2-l)\dot\alpha(l)}{}_{\beta(\hat{s}_1-k)\dot\beta(k)} 
$$
Recall also that
$$
\hat{s}_1 = s_2 + s_3 - s_1 - 1, \quad
\hat{s}_2 = s_1 + s_3 - s_2 - 1, \quad
\hat{s}_3 = s_1 + s_2 - s_3 - 1
$$
Now let us consider variations of the deformed curvatures
$\hat{\cal F} = {\cal F} + \Delta {\cal F}$ under the lowest spin 
$\omega$ gauge transformations. From the ansatz given above we can
immediately read the corrections to the gauge transformations:
\begin{equation}
\delta \Sigma^{\alpha(2s_1-2-m)\dot\alpha(m)} = a_{k,l,m}
\Omega^{\alpha(\hat{s}_3-m+l)\beta(\hat{s}_1-k)\dot\alpha(m-l)\dot\beta(k)}
\eta^{\alpha(\hat{s}_2-l)\dot\alpha(l)}{}_{\beta(\hat{s}_1-k)\dot\beta(k)} 
\end{equation}
Taking into account these corrections, the variation of the deformed
curvature $\hat{\cal F} = {\cal F} + \Delta {\cal F}$ appears to be
\begin{eqnarray}
\delta \hat{\cal F}^{\alpha(2s_1-2-m)\dot\alpha(m)} &=& a_{k,l,m} D
\Omega^{\alpha(\hat{s}_3-m+l)\beta(\hat{s}_1-k)\dot\alpha(m-l)\dot\beta(k)}
\eta^{\alpha(\hat{s}_2-l)\dot\alpha(l)}{}_{\beta(\hat{s}_1-k)\dot\beta(k)} \nonumber \\
  && + (a_{k,l,m+1} - a_{k,l-1,m}) e^\alpha{}_{\dot\gamma}
\Omega^{\alpha(\hat{s}_3-m+l-1)\beta(\hat{s}_1-k)\dot\alpha(m-l+1)\dot\beta(k)} \nonumber \\
 && \qquad \qquad 
\eta^{\alpha(\hat{s}_2-l)\dot\alpha(l-1)\dot\gamma}{}_{\beta(\hat{s}_1-k)\dot\beta(k)} \nonumber \\
  && + (k+1)a_{k+1,l,m} e^\gamma{}_{\dot\gamma}
\Omega^{\alpha(\hat{s}_3-m+l)\beta(\hat{s}_1-k-1)\dot\alpha(m-l)\dot\beta(k)\dot\gamma}  \eta^{\alpha(\hat{s}_2-l)\dot\alpha(l)}{}_{\beta(\hat{s}_1-k-1)\gamma\dot\beta(k)} \nonumber \\
 && + a_{k,l,m+1} e^\alpha{}_{\dot\gamma}
\Omega^{\alpha(\hat{s}_3-m+l-1)\beta(\hat{s}_1-k)\dot\alpha(m-l)\dot\beta(k)\dot\gamma}
\eta^{\alpha(\hat{s}_2-l)\dot\alpha(l)}{}_{\beta(\hat{s}_1-k)\dot\beta(k)} \nonumber \\
  && + (a_{k,l,m-1} - a_{k,l+1,m}) e_\gamma{}^{\dot\alpha}
\Omega^{\alpha(\hat{s}_3-m+l+1)\beta(\hat{s}_1-k)\dot\alpha(m-l-1)\dot\beta(k)} \nonumber \\
 && \qquad \qquad
\eta^{\alpha(\hat{s}_2-l-1)\gamma\dot\alpha(l)}{}_{\beta(\hat{s}_1-k)\dot\beta(k)} \nonumber \\
 && + a_{k,l,m-1} e_\gamma{}^{\dot\alpha}
\Omega^{\alpha(\hat{s}_3-m+l)\beta(\hat{s}_1-k)\gamma\dot\alpha(m-l-1)\dot\beta(k)}
\eta^{\alpha(\hat{s}_2-l)\dot\alpha(l)}{}_{\beta(\hat{s}_1-k)\dot\beta(k)} \nonumber \\
   && + (\hat{s}_1-k+1)a_{k-1,l,m} e_\gamma{}^{\dot\gamma}
\Omega^{\alpha(\hat{s}_3-m+l)\beta(\hat{s}_1-k)\gamma\dot\alpha(m-l)\dot\beta(k-1)} \nonumber \\
 && \qquad \qquad 
\eta^{\alpha(\hat{s}_2-l)\dot\alpha(l)}{}_{\beta(\hat{s}_1-k)\dot\beta(k-1)\dot\gamma} 
\end{eqnarray}
The main requirement here is that the deformed curvatures transform
covariantly, so we must have
\begin{eqnarray}
\delta \hat{\cal F}^{\alpha(2s_1-2-m)\dot\alpha(m)}
  &=& a_{k,l,m} 
{\cal R}^{\alpha(\hat{s}_3-m+l)\beta(\hat{s}_1-k)\dot\alpha(m-l)
\dot\beta(k)} 
\eta^{\alpha(\hat{s}_2-l)\dot\alpha(l)}{}_{\beta(\hat{s}_1-k)
\dot\beta(k)} \nonumber \\
&=& a_{k,l,m} [ D
\Omega^{\alpha(\hat{s}_3-m+l)\beta(\hat{s}_1-k)\dot\alpha(m-l)\dot\beta(k)} \eta^{\alpha(\hat{s}_2-l)\dot\alpha(l)}{}_{\beta(\hat{s}_1-k)\dot\beta(k)} \nonumber \\
 && \qquad + e^\alpha{}_{\dot\gamma}
\Omega^{\alpha(\hat{s}_3-m+l-1)\beta(\hat{s}_1-k)\dot\alpha(m-l)\dot\beta(k)\dot\gamma} \eta^{\alpha(\hat{s}_2-l)\dot\alpha(l)}{}_{\beta(\hat{s}_1-k)\dot\beta(k)} \nonumber \\
 && \qquad + (\hat{s}_1-k) e^\gamma{}_{\dot\gamma}
\Omega^{\alpha(\hat{s}_3-m+l)\beta(\hat{s}_1-k-1)\dot\alpha(m-l)\dot\beta(k)\dot\gamma} \eta^{\alpha(\hat{s}_2-l)\dot\alpha(l)}{}_{\beta(\hat{s}_1-k-1)\gamma\dot\beta(k)} \nonumber \\
 && \qquad + e_\gamma{}^{\dot\alpha}
\Omega^{\alpha(\hat{s}_3-m+l)\beta(\hat{s}_1-k)\gamma\dot\alpha(m-l-1)\dot\beta(k)} \eta^{\alpha(\hat{s}_2-l)\dot\alpha(l)}{}_{\beta(\hat{s}_1-k)\dot\beta(k)} \nonumber \\
 && \qquad + k e_\gamma{}^{\dot\gamma}
\Omega^{\alpha(\hat{s}_3-m+l)\beta(\hat{s}_1-k)\gamma\dot\alpha(m-l)\dot\beta(k-1)} \eta^{\alpha(\hat{s}_2-l)\dot\alpha(l)}{}_{\beta(\hat{s}_1-k)\dot\beta(k-1)\dot\gamma} ] 
\end{eqnarray}
A comparison of these two expressions gives us a number of recurrent
relations on the coefficients $a_{k,l,m}$
$$
a_{k,l,m+1} = a_{k,l-1,m}, \qquad a_{k,l,m-1} = a_{k,l+1,m}
$$
$$
(\hat{s}_1-k+1)a_{k-1,l,m} = ka_{k,l,m}, \qquad
(k+1)a_{k+1,l,m} = (\hat{s}_1-k)a_{k,l,m}
$$
$$
a_{k,l,m+1} = a_{k,l,m}, \qquad a_{k,l,m-1} = a_{k,l,m}
$$
their simple solution being
\begin{equation}
a_{k,l,m} = \frac{(\hat{s}_1)!}{(\hat{s}_1-k)!k!}a_0
\end{equation}
Thus the result turns out to be unique up to the one arbitrary
coupling constant.

\section{Flat limit}

The main problem with the flat limit is that the formalism we use
generates a lot of terms with the number of derivatives greater than
that of the flat vertex and their coefficients are singular in the
limit $\lambda \to 0$. Our first task here is to show that all such
terms combine into total derivatives or vanish on-shell and so they
all can be dropped out allowing us to take a desired limit. Let us
consider contribution to the cubic vertex from the highest spin field
as an example. They have the form (schematically)
$$
\Delta {\cal L}_1 \sim \sum_m {\cal F}_{\alpha(2s_1-2-m)\dot\alpha(m)}
\Delta {\cal F}^{\alpha(2s_1-2-m)\dot\alpha(m)}
$$
where $\Delta {\cal F}$ are given in (\ref{anz}). Recall that on-shell
each auxiliary or extra field 
$\Sigma^{\alpha(s_1-1+m_1)\dot\alpha(s_1-1-m_1)}$ 
is equivalent to $|m_1|$ derivatives of the physical one, in this the
number of derivatives for each concrete term in the cubic vertex is
defined by $N=|m_1| + |m_2| + |m_3| + 1$, where
\begin{eqnarray}
m_1 &=& s_1-1-m \nonumber \\
m_2 &=& s_2-1-m+l - k \label{par} \\
m_3 &=& s_3-1 - k - l \nonumber
\end{eqnarray}
Let us consider the contributions with positive $m_1$, while $m_{2,3}$
can be both positive or negative. Now we consider all four possible
cases, calculate the number of derivatives and focus on terms with
more than $N_0 = s_1+s_2-s_3$ derivatives.\\
I) $m_2>0$, $m_3>0$
$$
N = s_1+s_2+s_3-2 - 2m - 2k > s_1+s_2-s_3 \Rightarrow k < s_3-1-m
$$
II) $m_2>0$, $m_3<0$
$$
N = s_1+s_2 - s_3 - 2m + 2l > s_1+s_2-s_3 \Rightarrow l > m
$$
III) $m_2 < 0$, $m_3>0$
$$
N = s_1 - s_2 + s_3 - 2l > s_1+s_2-s_3 \Rightarrow l < s_3-s_2 <0
$$
IV) $m_2 <0$, $m_3<0$
$$
N = s_1-s_2-s_3 + 2 + 2k > s_1+s_2-s_3 \Rightarrow k > s_2-1 > 
\hat{s}_1
$$
So we see that only terms where all three $m_{1,2,3}$ are positive (or
all three are negative) generate the higher derivatives terms. Each
such contribution looks (schematically)
$$
[ D \Sigma - e\Sigma - \lambda^2e\Sigma] \Omega \omega
$$
so that we have terms with explicit derivative as well as the purely
algebraic ones. Let us begin with terms $D\Sigma\Omega \omega$.
Taking into account all combinatoric coefficients (both from the free
Lagrangian as well as from the deformation parameters) we obtain
\begin{equation}
\Delta =  C_{k,l,m}
 D \Sigma_{\alpha(\hat{s}_3-m+l)\delta(\hat{s}_2-l)\dot\alpha(m-l)
\dot\delta(l)} 
\Omega^{\alpha(\hat{s}_3-m+l)\beta(\hat{s}_1-k)\dot\alpha(m-l)\dot\beta(k)}
\omega^{\delta(\hat{s}_2-l)\dot\delta(l)}{}_{\beta(\hat{s}_1-k)
\dot\beta(k)}
\end{equation}
where
\begin{equation}
C_{k,l,m} = \frac{(\hat{s}_2+\hat{s}_3)!(\hat{s}_1)!a_0}
{(\hat{s}_3-m+l)!(\hat{s}_2-l)!(m-l)!l!(\hat{s}_1-k)!k!}
\end{equation}
Calculating the inverse relations from the (\ref{par})
\begin{equation}
m = (s_1-1) - m_1, \qquad
k = \frac{\hat{s}_1+\hat{m}_1}{2}, \qquad
l = \frac{\hat{s}_2+\hat{m}_2}{2}
\end{equation}
where we have introduced
\begin{equation}
\hat{m}_1 = m_1-m_2-m_3, \qquad
\hat{m}_2 = m_2-m_1-m_3, \qquad
\hat{m}_3 = m_3-m_1-m_2
\end{equation}
we can show that the denominator in the expression for $C_{k,l,m}$ can
be rewritten as follows:
$$
\left(\frac{\hat{s}_1+\hat{m}_1}{2}\right)!\left(\frac{\hat{s}_1-\hat{m}_1}{2}\right)!
\left(\frac{\hat{s}_2+\hat{m}_2}{2}\right)!\left(\frac{\hat{s}_2-\hat{m}_2}{2}\right)!
\left(\frac{\hat{s}_3+\hat{m}_3}{2}\right)!\left(\frac{\hat{s}_3-\hat{m}_3}{2}\right)!
$$
Taking into account the relations on the constants $a_0$, $b_0$ and
$c_0$, we see that such contributions are completely symmetric on the
three fields. As a result, all such terms with explicit derivative
combine into total derivative exactly in the same way as the terms
with the highest number of derivatives do. 

Now we consider purely algebraic terms of the type 
$\lambda^2e\Sigma\Omega\omega$. We obtain
\begin{eqnarray}
\Delta_1 &=& (\hat{s}_3-m+l)C_{k,l,m} e_\gamma{}^{\dot\gamma}
\Sigma_{\alpha(\hat{s}_3-m+l-1)\delta(\hat{s}_2-l)
\dot\alpha(m-l)\dot\delta(l)\dot\gamma} \nonumber \\
 && \Omega^{\alpha(\hat{s}_3-m+l-1)\beta(\hat{s}_1-k)\gamma
\dot\alpha(m-l)\dot\beta(k)}
\omega^{\delta(\hat{s}_2-l)\dot\delta(l)}{}_{\beta(\hat{s}_1-k)
\dot\beta(k)} + \dots
\end{eqnarray}
where dots stand for the similar terms with index $\gamma$ contracted
with one of the indices of the field $\omega$. On the other hand, if
we take the contribution of the type $e\Omega\Sigma\omega$ from the
deformations of the $\Omega$ field, we obtain
\begin{eqnarray}
\Delta_2 &=& (\tilde{m}-\tilde{l}) 
\tilde{C}_{\tilde{k},\tilde{l},\tilde{m}}
 e^\gamma{}_{\dot\gamma} \Omega_{\alpha(\hat{s}_3-\tilde{m}+\tilde{l})
\delta(\hat{s}_1-\tilde{l})\dot\alpha(\tilde{m}-\tilde{l}-1) 
\dot\delta(\tilde{l})} \nonumber \\
 && \Sigma^{\alpha(\hat{s}_3-\tilde{m}+\tilde{l})
\beta(\hat{s}_2-\tilde{k})\dot\alpha(\tilde{m}-\tilde{l}-1)\dot\beta(\tilde{k})}
\omega^{\delta(\hat{s}_1-\tilde{l})\dot\delta(\tilde{l})}{}_{
\beta(\hat{s}_2-\tilde{k})\dot\beta(\tilde{k})} + \dots
\end{eqnarray}
where again dots stand for the similar terms where index $\dot\gamma$
is contracted with one of the $\omega$ indices. We see that the
structure of these two contributions is the same provided
\begin{equation}
\tilde{m} = m-l+k+1, \qquad \tilde{k} = l, \qquad \tilde{l} = k
\end{equation}
The resulting coefficients turn out to be equal so these two terms
cancel each other. The same holds for the two other pairs of
contractions, namely $(\Sigma\omega)$ and $(\Omega\omega)$.

Thus all the higher derivative terms combine into total derivatives or
cancel each other and we may safely take a flat limit.
We repeat our considerations but focus this time on the terms with 
exactly $N_0=s_1+s_2-s_3$ derivatives, i.e. those which do not vanish 
in the flat limit. We find that there are a lot of such terms with
both positive and negative $m_{2,3}$. The situation with positive
$m_{2,3}$ appears to be mainly the same as before, so that they also
combine into total derivatives or cancel. As for the terms with
negative $m_2$ or/and $m_3$, after rather long work we have managed to
show that most of them can be combined into terms proportional to the
gauge invariant curvatures which vanish on-shell. All this leads to
the surprisingly simple results presented in the main text.

\end{document}